\documentclass{aa}
\usepackage{times,float}
\usepackage{epsfig}
\usepackage{graphics}
\usepackage{amssymb}
\usepackage{float}

\def\tj{\theta_{\rm{jet}}}
\def\to{\theta_{\rm{obs}}}
\def\tio{t_{\rm{obs}}}
\def\tij{t_{\rm{jet}}}
\def\ltsima{$\; \buildrel < \over \sim \;$}
\def\lsim{\lower.5ex\hbox{\ltsima}}
\def\gtsima{$\; \buildrel > \over \sim \;$}
\def\gsim{\lower.5ex\hbox{\gtsima}}

\begin{document}

\title{On the jet structure and magnetic field configuration of
GRB~020813\thanks{Based on observations collected at the European
Southern Observatory, Cerro Paranal (Chile), ESO programmes
69.D-0461(A) and 69.D-0701(A).}}

\titlerunning{Jet structure of GRB~020813}
\author{D.~Lazzati\inst{1}
       \and S.~Covino\inst{2}
       \and J.~Gorosabel\inst{3,4}
       \and E.~Rossi\inst{1}
       \and G.~Ghisellini\inst{2}
       \and E.~Rol\inst{5} 
       \and J.~M.~Castro Cer\'on\inst{4}
       \and A.~J.~Castro-Tirado\inst{3}
       \and M.~Della~Valle\inst{6}
       \and S.~di~Serego~Alighieri\inst{6}
       \and A.~S.~Fruchter\inst{4}
       \and J.~P.~U.~Fynbo\inst{8}
       \and P.~Goldoni\inst{9}
       \and J.~Hjorth\inst{10} 
       \and G.~L.~Israel\inst{7}
       \and L.~Kaper\inst{5}
       \and N.~Kawai\inst{11} 
       \and E.~Le~Floc'h\inst{9}
       \and D.~Malesani\inst{12}
       \and N.~Masetti\inst{13}
       \and P.~Mazzali\inst{14,15,16}
       \and F.~Mirabel\inst{9}
       \and P.~M\o{}ller\inst{17}
       \and S.~Ortolani\inst{18}
       \and E.~Palazzi\inst{13}
       \and E.~Pian\inst{14}
       \and J.~Rhoads\inst{4}
       \and G. Ricker\inst{19} 
       \and J.~D.~Salmonson\inst{20}
       \and L.~Stella\inst{7}
       \and G.~Tagliaferri\inst{2}  
       \and N.~Tanvir\inst{21}
       \and E.~van~den~Heuvel\inst{5}
       \and R.~A.~M.~J.~Wijers\inst{5}
       \and F.~M.~Zerbi\inst{2}}
\institute{
Institute of Astronomy, University of Cambridge, Madingley Road,
UK--CB3 0HA Cambridge, UK
\and
INAF, Osservatorio Astronomico di Brera, via E. Bianchi 46, I--23807
Merate (LC), Italy
\and
Instituto de Astrof\'{\i}sica de Andaluc\'{\i}a (IAA-CSIC), Apartado
de Correos, 3.004, E-18.080 Granada, Spain
\and 
Space Telescope Science Institute, 3700 San Martin Drive, Baltimore,
MD 21218-2463, USA
\and
University of Amsterdam, Kruislaan 403, NL--1098 SJ Amsterdam, The
Netherlands
\and
INAF, Osservatorio Astrofisico di Arcetri, Large E. Fermi 5, I--50125
Firenze, Italy
\and
INAF, Osservatorio Astronomico di Roma, via Frascati 33, I--00044
Monterporzio, Italy
\and
Department of Physics and Astronomy, University of \AA rhus, Ny
Munkegade, DK--8000 \AA rhus C, Denmark
\and
CEA/DSM/DAPNIA, L'Orme des Merisiers, Bat. 709, F--91191
Gif-sur-Yvette, France
\and
Niels Bohr Institute, Astronomical Observatory, University of
Copenhagen, Juliane Maries Vej 30, DK--2100 Copenhagen \O, Denmark
\and
Department of Physics, Faculty of Science, Tokyo Institute of
Technology 2-12-1 Ookayama, Meguro-ku, Tokyo 152-8551, Japan
\and
International  School for  Advanced Studies  (SISSA/ISAS), via  Beirut 2-4,
I--34014 Trieste, Italy
\and
Istituto di Astrofisica Spaziale e Fisica Cosmica - Sezione di Bologna, 
CNR, Via Gobetti 101, I-40129 Bologna, Italy
\and
INAF, Osservatorio Astronomico di Trieste, Via Tiepolo 11,
I-34131 Trieste, Italy
\and 
Department of Astronomy, University of Tokyo, Bunkyo-ku, Tokyo 
113-0033, Japan
\and
Research Center for the Early Universe, University of
Tokyo, Bunkyo-ku, Tokyo 113-0033, Japan
\and
European Southern Observatory, Karl--Schwarzschild--Stra\ss e 2,
D--85748 Garching, Germany
\and
Universit\`a di Padova, Dip. di Astronomia, Vicolo dell'Osservatorio
2, I--35122 Padova, Italy
\and
Center   for  Space  Research,   Massachusetts  Institute   of  Technology,
Cambridge, MA 02139-4307, USA
\and
Lawrence Livermore National laboratory, L-038, P.O. Box 808,
Livermore, CA, 94551
\and
Department of Physical Sciences, University of Hertfordshire, College
Lane, Hatfield, Herts UK--AL10 9AB, UK
}
\offprints{D. Lazzati, {\tt lazzati@ast.cam.ac.uk}}

\date{Received / Accepted }

\abstract{The polarization curve of GRB~020813 is discussed and
compared to different models for the structure, evolution and
magnetisation properties of the jet and the interstellar medium onto
which the fireball impacts. GRB~020813 is best suited for this kind of
analysis for the smoothness of its afterglow light curve, ensuring the
applicability of current models. The polarization dataset allows us to
rule out the standard GRB jet, in which the energy and Lorentz factor
have a well defined value inside the jet opening angle and the
magnetic field is generated at the shock front. We explore alternative
models finding that a structured jet or a jet with a toroidal
component of the magnetic field can fit equally well the polarization
curve. Stronger conclusions cannot be drawn due to the incomplete
sampling of the polarization curve. A more dense sampling, especially
at early times, is required to pin down the structure of the jet and
the geometry of its magnetic field.
}

\maketitle

\section{Introduction}

The discovery that Gamma-Ray Burst (hereafter GRB) afterglows are
linearly polarized is one of the strongest proofs that the photons we
observe are synchrotron radiation (Covino et al. 1999; Wijers et
al. 1999). Despite that, polarimetric afterglow observations have so
far been sparse and discontinuous, with only few polarimetric
measurements performed on different optical transients (see Covino et
al. 2003b for a review). Recently, however, several better sampled linear
polarization measurements allowed for the first studies
of the evolution of polarization, especially for GRB~021004 (Rol et
al. 2003; Lazzati et al. 2003) and GRB~030329 (Greiner et al. 2003).

These studies (Lazzati et al. 2003; Nakar \& Oren 2003), coupled with
theoretical works (Ghisellini \& Lazzati 1999; Sari 1999; Rossi et
al. 2004; Granot \& K\"onigl 2003) reached two important
conclusions. First, polarimetric studies provide unique information on
the structure and dynamics of GRB outflows: the polarization, e.g.,
from a homogeneous jet has a markedly different evolution from that of
a structured one, even though their light-curves are barely
distinguishable (Rossi et al. 2004). Second, polarization studies are
complex and subject to systematic errors. This is due to the
combination of the intrinsic weakness of GRB polarization ($\lsim
3\%$, at least at the times at which measurements have so far been
possible; Covino et al. 2003b, but see also Bersier et al. 2003 who
found a possible polarization flickering at the 10\% level) and to the
sensitivity of the polarization signal to bright spots on the fireball
surface and/or inhomogeneities in the ambient medium (Lazzati et
al. 2003; Granot \& K\"onigl 2003; Nakar \& Oren 2003). In addition,
the measured polarized signal is comparable to the polarization
induced by the propagation of light through a moderately absorbing
interstellar material, making the direct comparison of models with
data more difficult (Lazzati et al. 2003).

All these considerations make the dataset obtained for GRB~020813
unique and extremely interesting. GRB~020813 (see the accompanying
paper, Gorosabel et al. 2003, for more details) had an extremely
smooth light curve (Laursen \& Stanek 2003; Gorosabel et al. 2003),
successfully fitted by a smoothly broken power-law with an
r.m.s. scatter of the residuals of $<0.01$ magnitudes in the optical
filters. This ensures that inhomogeneities in the fireball structure
and/or in the surrounding interstellar medium (ISM) are not
significant and therefore can not affect the polarization
measurement. In addition, the spectropolarimetric measurement of Barth
et al. (2003) does not show evidence of a strong colour dependence of
the polarization, a signature of the polarization induced by the
interstellar medium (Serkowski et al. 1975).

In this paper we compare the polarization evolution of GRB~020813 with
existing models from the production of polarized light in spherical
and beamed fireballs. Some of these models, which were originally
computed in a uniform environment, are extended to the wind case
(Sect.~2). We also consider the possible presence of an ordered
component of the magnetic field advected from the central source
(Sect. 2). The comparison of the models with the data is described in
Sect.~3 and we finally discuss our findings in Sect.~4.

\section{The models}

In this section we describe the models we will compare to the
polarimetric data of GRB~020813. Although some of the theoretical
polarization curves are taken from the literature those associated
with magnetised jets are originally calculated for this work.

First, let us consider the magnetic domain model, the first ever
considered model for the observation of linear polarization in GRB
afterglows (Gruzinov \& Waxman 1999). If the shock generated field is
able to rearrange rapidly in ordered domains, Gruzinov \& Waxman
(1999) calculated that an average observer should see about $~\sim50$
magnetic domains, and therefore if each domain produces a polarization
$p_0$ a net polarization
$\Pi=p_0/\sqrt{N}\sim0.1(p_0/0.7)(50/N)^{1/2}$ should be observed,
where $p_0\sim70\%$ for synchrotron. Deriving a polarization curve for
this model is not an easy task, given its intrinsic random
character. Nevertheless, a general conclusion can be
drawn. Polarization should be variable and variability in the degree
of polarization should be associated to variability in the position
angle (Gruzinov \& Waxman 1999). In particular, a variation of linear
polarization by a factor of 2 should be associated to a random
re-shuffling of the position angle. This is not what we observe in the
data, where only a moderate rotation of a few degrees is associated to
the polarization evolution (Gorosabel et al. 2003). We conclude
therefore, analogously to what derived by Barth et al. (2003) (see
also Greiner at al. 2003 for the case of GRB~030329), that random
patches of ordered magnetic field are not producing the observed
polarization.

\begin{figure}
\psfig{file=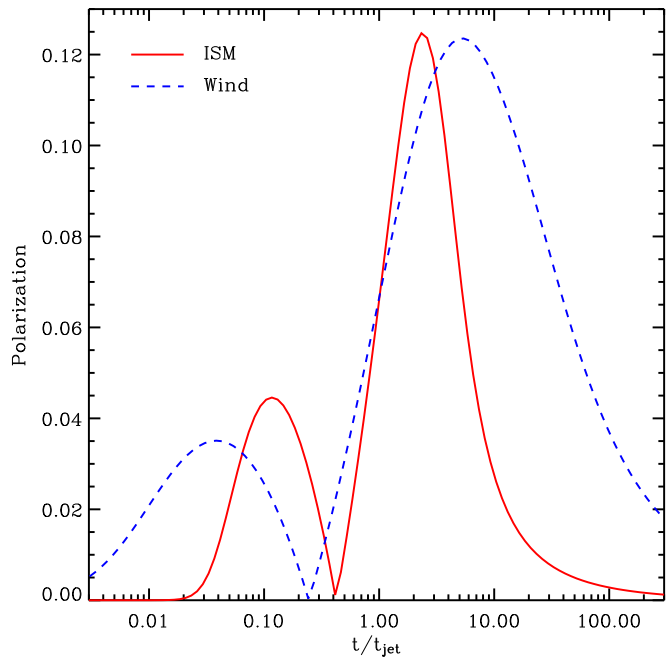,width=\columnwidth}
\caption{{Polarization curves from a homogeneous non sideways expanding 
jet with shock generated magnetic field in a standard ISM and wind
environment. Both curves are computed for observers located at
$\theta_{\rm{obs}}=0.67\theta_{\rm{jet}}$. The break time
$t_{\rm{jet}}$ has been measured independently for the two light
curves. The polarization position angle rotates by $90\degr$ in the
moment of null polarization.}
\label{fig:ismw}}
\end{figure}

\subsection{Hydrodynamic Homogeneous Jet}

\begin{figure}
\psfig{file=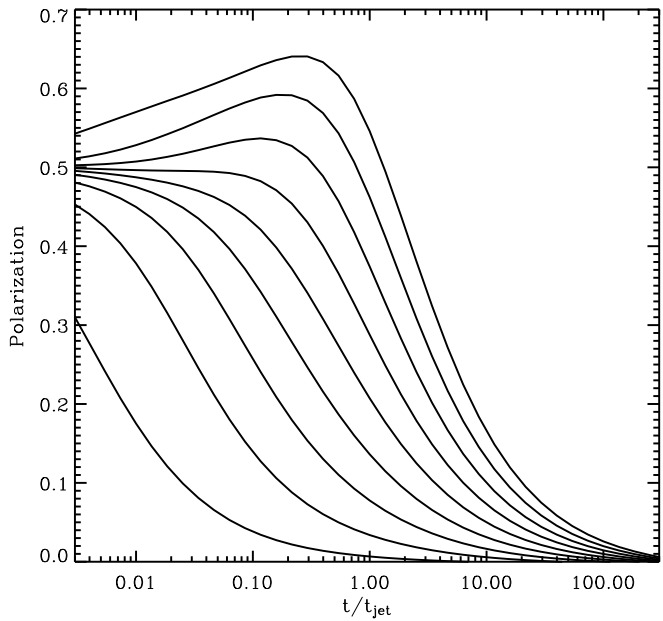,width=\columnwidth}
\caption{{Polarization curves for an MHJ jet with toroidal magnetic
field. From left to right different curves refer to $\to/\tj=0.1$,
0.2, 0.3, 0.4, 0.5, 0.6, 0.7, 0.8 and 0.9. The position angle of the
polarization is constant throughout the entire evolution.}
\label{fig:lyu1}}
\end{figure}

We define a ``Hydrodynamic Homogeneous Jet'' (HHJ) as a standard
top-hat jet in which the energy, which is uniformly distributed within
the jet opening angle, is carried by baryons and the magnetic field
responsible for the afterglow synchrotron emission is tangled on small
scales but overall dominated by either a component orthogonal or
parallel to the shock front. Such a field configuration is generated
either by compression of a fully tangled magnetic field of by two
stream plasma instabilities at the shock front, for example the Weibel
instability (Silva et al. 2003). Polarization curves from this class
of jets have been computed by various authors (Ghisellini
\& Lazzati 1999; Sari 1999; Granot \& K\"onigl 2003; Salmonson 2003;
Rossi et al. 2004). Most of these papers, and in particular the more
recent ones based on numerical computations rather than on analytical
approximations, agree qualitatively on the resulting polarization
curve: it has two peaks, separated by a moment of null polarization
roughly coincident with the break time of the total light curve. In
this moment the position angle of the polarization rotates by
$90\degr$. The second peak is always stronger than the first, their
ratio depending on the dynamics of the jet sideways expansion (SE):
the faster the expansion, the smaller the second peak (the first peak
is obviously only marginally affected by sideways expansion; see Rossi
et al. 2004). In this paper we adopt the polarization curves
computed by Rossi et al. (2004). After the jet break time we consider
either no lateral expansion or a jet expanding at the speed of sound
in the shocked fluid comoving frame (Eq. 8 of Rossi et
al. 2004). These two assumptions are chosen in order to encompass the
numerical results of Kumar \& Granot (2003), who find a sub-sonic
expansion until the expansion velocity becomes
trans-relativistic. Extremely high sideways expanding efficiency were
instead assumed by Sari (1999). In this case a different behaviour of
the post-break polarization is obtained, where the polarization can
have either no change in position angle or a double change (see also
Barth et al. 2003). The energy density within the jet opening angle is
assumed to be uniform before and after the break time.

In addition to the calculations presented in the above mentioned
papers, we show here the effect of a wind environment on the
polarization curve. In Fig.~\ref{fig:ismw} a polarization curve for a
HHJ expanding in an ISM environment (solid line) is compared to that
of the very same jet that propagates in a wind (where
$n(r)\propto{}r^{-2}$; dashed line). The qualitative result is
identical, with the polarization curve characterised by two peaks with
position angle shifted by $90\degr$. The only difference is that in
the wind case the evolution is slower, as already noted by Kumar \&
Panaitescu (2000) for the break of the total light curve. Before the
non-relativistic transition ($t\ll{}t_{\rm{NR}}$) it is possible to
obtain the wind polarization curve with good approximation from the
ISM one rescaling the times according to
$t_{\rm{wind}}=t_{\rm{ISM}}^{3/2}$. This, for a non sideways expanding
jet, comes from the fact that the polarization behaviour mainly
depends on the Lorentz factor $\Gamma$ which scales as $t^{-3/8}$ for
the ISM case and as $t^{-1/4}$ in the wind environment.

\subsection{Hydrodynamic Structured Jet}

A ``Hydrodynamic Structured Jet'' (HSJ) is similar to a HHJ from the
micro-physical point of view, but has a distribution of energy per
unit solid angle which is larger in the centre and smaller in the
wings (Rossi et al. 2002). In order to preserve the standard jet
energy (Frail et al. 2001), this distribution must have the form of a
power-law with the energy scaling as $\theta^{-2}$. Polarization
curves for this jet have been computed by Rossi et al. (2002;
2004). They are single peaked, with the time of the maximum coincident
with the break time of the total light curve. The position angle does
not vary throughout the evolution, and the maximum observed
polarization grows with the angle that the line of sight makes with
the jet axis. For this case we did not consider sideways
expansion since it affects only marginally the behaviour of the
polarization curve, to a level much smaller than the accuracy of the
data we are dealing with. A discussion of theoretical light curves can
be found in Rossi et al. (2004).

\subsection{Magnetised Homogeneous Jet}

\begin{figure}
\psfig{file=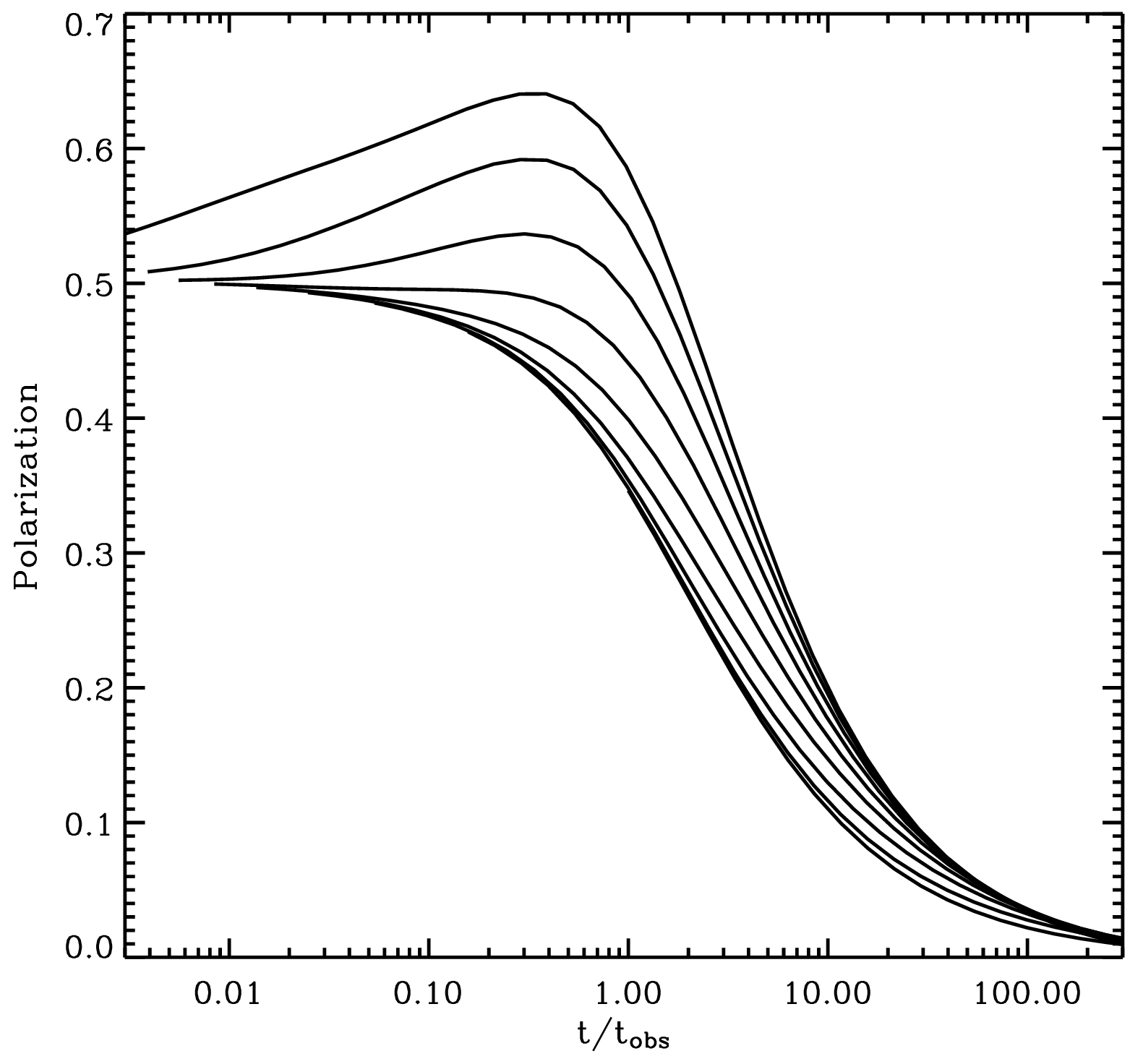,width=\columnwidth}
\caption{{Same as Fig.~\ref{fig:lyu1} but with the time shown in units
of $\tio$ (defined through $\Gamma(\tio)=\to^{-1}$) rather than $\tij$
(defined through $\Gamma(\tij)=\tj^{-1}$). }
\label{fig:lyu2}}
\end{figure}

We call here ``Magnetised Homogeneous Jet'' (MHJ) a jet in which the
magnetic field has a toroidal structure, with the polar axis
coincident with the jet axis. The energy is uniform within the jet
opening angle, analogously to a HHJ. Such a magnetic configuration can
be realized, e.g., if the energy is carried in rough equipartition by
the baryons and an electromagnetic component (e.g. Proga et
al. 2003). Such a scenario is naturally envisaged in the original GRB
jet, where a sizable magnetic field component can be advected from the
central engine. Whether this field can be propagated to the external
shock is less certain, but such a possibility should be considered
(Lyutikov, Pariev \& Blandford 2003; Lyutikov \& Blandford 2004),
given the fact that contact discontinuities are known to be maximally
unstable (Landau \& Lifshitz 1989) due to the lack of any restoring
force. Here we assume that all the synchrotron emission is produced by
the toroidal magnetic field, ignoring a possible (even likely, given
the low measured levels of polarization) random component of the field
(see below). We also consider uniform jets, i.e. jets with a
well-defined cone angle within which the jet is uniform. Structured
magnetised jets will be discussed in the next section. Polarization
curves are shown in Figs.~\ref{fig:lyu1} and~\ref{fig:lyu2}.

There are clearly major differences with respect to the hydrodynamic
jets (either homogeneous or not). The first is that the polarization
does not disappear for very small observer times. This is due to the
fact that in the hydrodynamic jets the polarization is cancelled by
symmetry unless the edge of the jet (for the HHJ) or the brighter core
(HSJ) are visible. In the MHJ case, the observed afterglow
polarization is not due to a lack of cancellation, but rather to the
presence of a genuinely ordered magnetic field. The higher the Lorentz
factor of the fireball (and therefore the smaller the observer time)
the less curved is the observed field on the plane of the sky and
therefore the higher the polarization (see e.g. Lyutikov et
al. 2003). For $\to/\tj<0.6$ the polarization curve is monotonically
decreasing, while for $\to/\tj>0.6$ there is a maximum in the
polarization curve at $t\sim\tio$, where $\tio$ is defined through
$\Gamma(\tio)=1/\to$. The position angle is constant throughout
the whole evolution.

Another important difference between hydrodynamic polarization curves
and MHJ ones is that in the case of hydrodynamic light and
polarization-curves there is only one relevant time-scale, while MHJ
curves have two time-scales. As can be seen by comparing
Fig.~\ref{fig:lyu1} with Fig.~\ref{fig:lyu2}, the break in the
polarization curve takes place approximately at $t=\tio$, while the
break in the total light curve happens at $t=\tij$. In the case of
hydrodynamic jets, the relevant change of behaviour in the
polarization curve (the change of angle in HHJs or the peak in HSJs)
takes place in coincidence with the change of slope in the total light
curve, i.e. the break time $\tij$. As a consequence, in an MHJ jet the
simultaneous observation of light and polarization-curves can
completely solve the jet and the observed geometry, something that is
impossible in a hydrodynamic jet.

Before discussing the last configuration, that of a force-free
magnetic bubble, it is worth stressing that the curves shown in
Figs.~\ref{fig:lyu1} and~\ref{fig:lyu2} heavily rely on the assumption
that the magnetic field contained in the ejecta is toroidal.
Such a field should be transported out to the shocked ISM producing
the afterglow radiation. In our computations it is assumed that all
the magnetic field present in the shocked ISM comes from the GRB
ejecta. It is more likely that a turbulent magnetic field is generated
at the shock and mixed with the ordered magnetic field of the
ejecta. This is also required by the fact that the typical afterglow
polarization is at the level of few per cent, much smaller than the
values predicted in Figs.~\ref{fig:lyu1} and~\ref{fig:lyu2}. The
simplest assumption that can be made is that the ratio between the
turbulent and ordered components of the field stays constant
throughout the evolution. In this case the resulting polarization
curve may be obtained by rescaling those in the figures. However, the
ratio between the two fields may change with time and in that case,
even if it is likely that the general monotonic behaviour would be
maintained, the actual shape of the curves may change. Given the
quality of the dataset and the theoretical difficulties in the
propagation of the field, we adopt a completely ordered field or a mix
with constant ratio.

\begin{figure}
\psfig{file=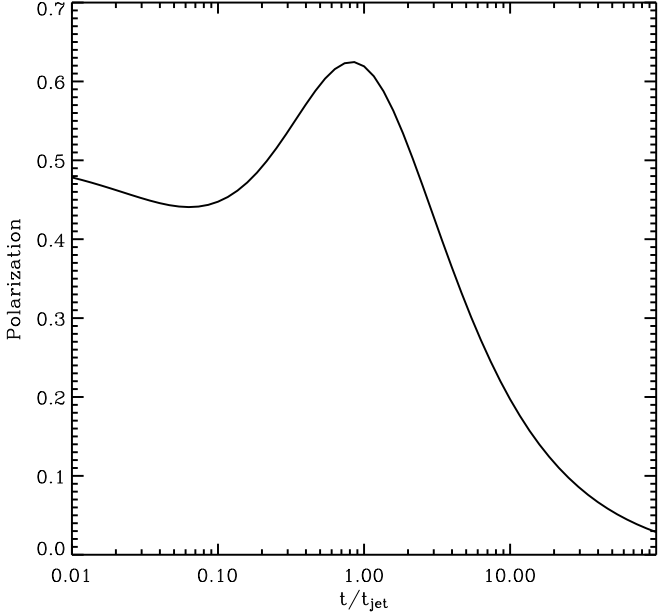,width=\columnwidth}
\caption{{Polarization curve for a MSJ with purely toroidal magnetic
field. The polarization curve is unique and independent of the
observer line of sight as long as the time is plotted in unit of the
light curve break time. The position angle of the polarization is
constant throughout the entire evolution.}
\label{fig:toro}}
\end{figure}

\subsection{Magnetised Structured Jet}

It has been proposed recently that many astrophysical jets may be
dominated by electromagnetic forms of energy rather than by baryonic
matter (Lyutikov \& Blandford 2004; Lyutikov et al. 2003). If the
plasma is sufficiently tenuous, the flow can be followed under the
Force-Free approximation, a set of equations in which the inertia of
the matter is neglected. We here consider the late phase of the
evolution, when the external shock has developed and its dynamics does
not differ any more from that of an hydrodynamic jet. Analogously to
the MHJ jet, we assume that the magnetic field in the shocked ISM is
transported from the magnetic bubble. We call this configuration a
``Magnetised Structured Jet'' (MSJ), i.e. a jet in which the
energy distribution is inhomogeneous as described for the HSJ, and the
magnetic field is toroidal as in the case of a MHJ. The difference
with respect to the MHJ jet is therefore that the jet is not uniform
within its opening angle but structured, with an energy per unit solid
angle distribution $E_\Omega\propto1/\sin^2\theta\sim\theta^{-2}$,
analogously to the HSJ. The polarization curve has been computed
From Eq. 2, 6 and 7 of Lyutikov et al. (2003) and is shown in
Fig.~\ref{fig:toro}. The polarization curve for this class of jets is
unique, independent of the observer line of sight $\to$, if plotted
against the observer time in units of the jet break time $\tij$. The
position angle is independent of time, since reflects the orientation
of the magnetic field. As for the MHJ jet, a random magnetic field
component is likely to be mixed in the toroidal field altering the
detailed shape of the light curve, but not its general behaviour.

\section{Comparison of the data with the models}

\begin{figure*}[t]
\parbox{\columnwidth}{\psfig{file=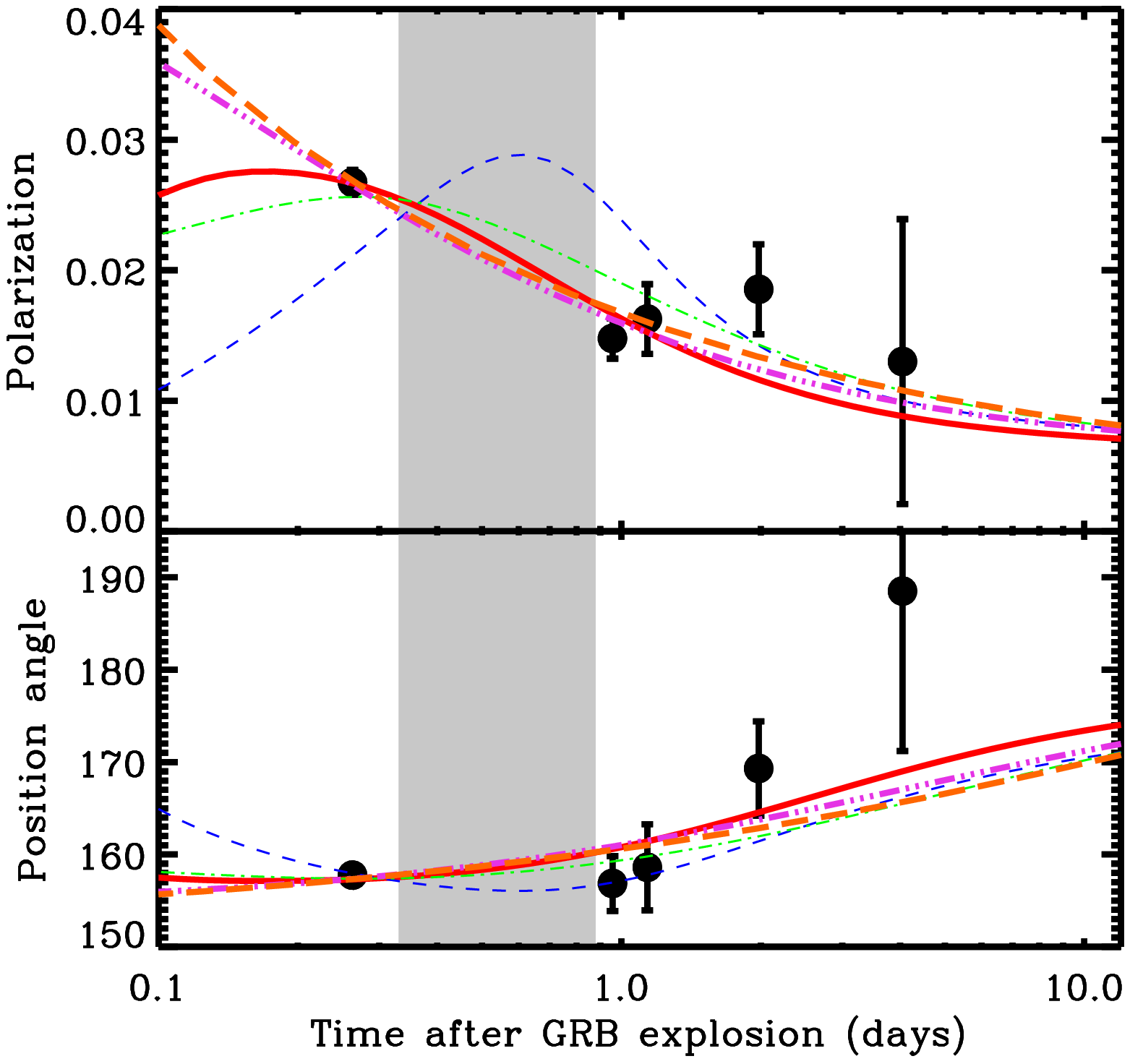,width=\columnwidth}
\caption{{Polarization (upper panel) and position angle (lower panel) 
data for GRB~020813 (Gorosabel et al. 2003). Different curves refer to
different models, as indicated in Fig.~\ref{fig:fit2}. Models yielding
an acceptable fit are plotted with a thick line, while non-acceptable
models are shown with a thin line. The gray-shaded area shows the
acceptable range for the jet break time.}
\label{fig:fit}}}
\hspace{3truemm}
\parbox{\columnwidth}{\psfig{file=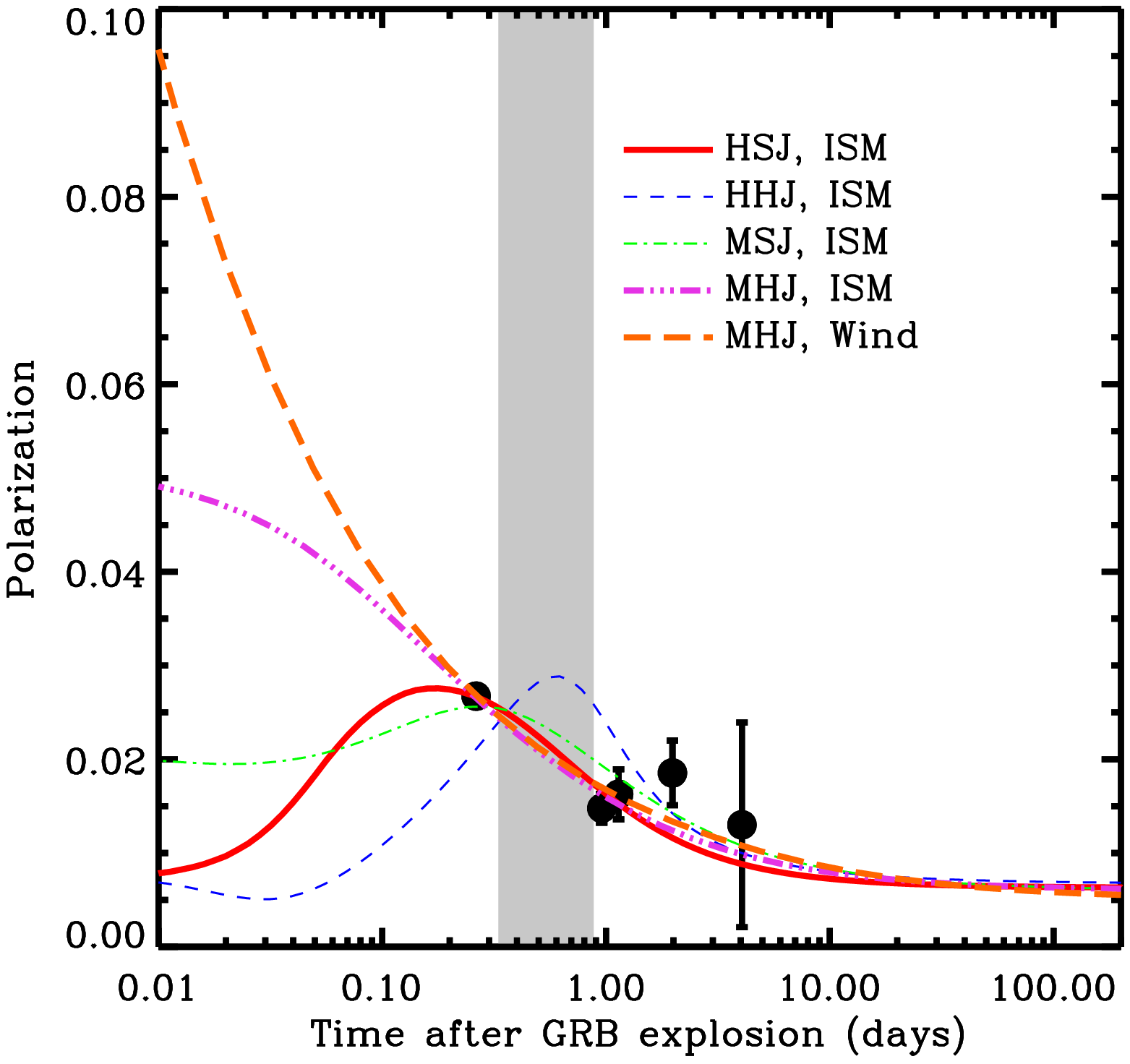,width=\columnwidth}
\caption{{Same as Fig.~\ref{fig:fit} but for the polarization only.
The $x$ and $y$ axes have been expanded in order to show how the
models differ at early times, where more data are needed in order to
clearly understand the structure of the jet and of the magnetic
field. The gray-shaded area shows the acceptable range for the jet
break time (Gorosabel et al. 2003).}
\label{fig:fit2}}}
\end{figure*}

\begin{table*}
\caption{{Fit results.}
\label{tab:fit}}
\begin{center}
\begin{tabular}{c|lllll}
Model$^{(a)}$& $b^{(b)}$     & $\tij^{(c)}$ & $\to/\tj$ & $\theta_{\rm{OT}}^{(d)}$ & $\chi^2/$d.o.f. \\ 
\hline
HHJ ISM     & $0.35$       & $0.33$       & $0.8$     & 151.0 & 84.7/6  \\
HHJ Wind    & $0.32$       & $0.33$       & $0.8$      & 151.0 & 119.2/6 \\
HHJ ISM SE  & $0.63$       & $0.33$       & $0.5$      & & 194.0/6 \\
HHJ Wind SE & $0.57$       & $0.33$       & $0.5$      & 150.0 & 194.2/6   \\
{\bf HSJ ISM}& $0.475\pm0.02$ & $0.29\pm0.07$ & $2.0\pm0.1$ & $152.0\pm1.2$ & 10.36/6 \\
HSJ Wind    & $0.46$       & $0.33$       & $2.0$      & & 18.7/6  \\
{\bf MHJ ISM}& $<0.6^{(e)}$  & $0.7^{(e)}$   & $0.4^{(e)}$ & $152.0\pm1.2$ &  8.9/6  \\
{\bf MHJ Wind}& $>0.5^{(e)}$ & $0.38^{(e)}$  & $0.13^{(e)}$& $152.0\pm1.2$ & 8.7/6   \\
MSJ ISM     & $0.185$      & $0.33$       &           & 152.0 & 15.8/7  \\
MSJ Wind    & $0.18$       & $0.33$       &           & 152.0 & 29.3/7  \\ \hline
\end{tabular}
\end{center}

\noindent
$^{(a)}$ Acronym of the considered model. Models yielding an acceptable
fit have been highlighted in bold. Models with SE have sideways expansion 
included in the computation.\\
$^{(b)}$ Alignment of the magnetic field (see text). \\
$^{(c)}$ The jet break time (in days) is constrained to be consistent 
with what 
measured from the light-curve (Gorosabel et al. 2003). In most cases 
the break time is forced to be the smallest possible (0.33 days). \\
$^{(d)}$ Position angle of the intrinsic OT polarization before the
propagation in the ISM. \\
$^{(e)}$ Despite the goodness of the fit it was impossible to derive 
meaningful error intervals for these parameters (see text).
\end{table*}

The polarization dataset for GRB~020813 has been presented elsewhere
(Gorosabel et al. 2003) and we here simply note two differences
between their data and the data we adopt. First, we adopt the dataset
uncorrected for polarization induced by the propagation into the
Galactic ISM, as derived from the polarization properties of field
stars. This choice is due to the fact that on the one hand the path of
the OT light in the ISM is longer than that of field stars, and
therefore the induced polarization may be different, and on the other
hand the host galaxy ISM may also contribute to the induced
polarization. Since the total induced polarization can be dealt with
using a single set of $q$ and $u$ Stokes parameters, we prefer to use
here the uncorrected data (as previously done with GRB~021004, Lazzati
et al. 2003), bearing in mind that the average Stokes parameters of
field stars are $q_{\rm{ISM}}=6.22\times10^{-3}$ and
$u_{\rm{ISM}}=-3.95\times10^{-4}$ (Gorosabel et al. 2003). The second
difference with respect to the dataset of Gorosabel et al. (2003) is
that the observations have been binned in time in order to increase
the signal-to-noise\footnote{Polarization points that have been binned
were consistent with each other.}. In particular, the first 4 data
points (with 750 s exposure each) in Tab.~1 of Gorosabel et al. (2003)
have been averaged into a single point, as well as the remaining 3
points of the first observing night (300 s exposure each). Our dataset
is presented in Fig.~\ref{fig:fit}. Even though we present
polarization and position angle data, the fits were performed in the
Stokes parameter space, where the uncertainties have a Gaussian
distribution.

Fit results are reported in Table~\ref{tab:fit} for all the models
described above. All the models have been fitted for evolution in a
uniform ISM as well as for a wind environment. In the case of HHJ
models, the possibility that the jet undergoes sideways expansion at
the internal sound speed was considered as well. In addition to the
jet parameters, the possible contribution of a polarizing ISM with free
properties was considered. The resulting fit was always consistent
with the $q_{\rm{ISM}}$ and $u_{\rm{ISM}}$ derived from the field
stars. We have therefore frozen the induced polarization to the
Galactic value in all the fits, in order to increase the number of
degrees of freedom. This result is also consistent with the low
upper-limit for reddening derived by Covino et al. (2003a) from
$UBVRIJHK$ quasi-simultaneous photometry and with the
spectropolarimetry of Barth et al. (2003). In order to check also for
the possible presence of an external ordered magnetic field (Granot \&
K\"onigl 2003) we have allowed the ISM Stokes parameters to become
comparable to the measured polarization. This did not lead to a
significant improvement of the fit.

Fits with a hydrodynamic homogeneous jet yield always non-acceptable
$\chi^2$ values. This is mainly due to the fact that a minimum of
polarization in coincidence with the break time is not observed, nor a
$90\degr$ rotation of the position angle between the data taken across
the jet break time. A reasonable (even though still not formally
acceptable) fit can be obtained only in two limiting cases, which are
not likely. If the jet break time is allowed to become smaller than
0.33~d, which is not consistent with the break time detected in the
light-curve (Gorosabel et al. 2003) or if the line of sight is allowed
to be slightly outside the jet edge in a non sideways expanding beam,
the measured polarization points lie in a time interval where the
angle rotation is not expected. In this case reasonably good fits can
be obtained (with reduced $\chi^2$ values of $\sim2\div3$). The best
HHJ fit is shown with a thin dashed line in Figs.~\ref{fig:fit}
and~\ref{fig:fit2}.

The fit with a structured hydrodynamic jet gives better results. This
is mainly due to the fact that in this case the position angle of
polarization is constant, and the polarization curve has a maximum in
coincidence with the break time. In this case a good fit is obtained
($\chi^2/$d.o.f.=10.36/6) for a jet expanding in a uniform medium
observed very near to its core, consistent with the small measured
opening angle from the light-curve break time (Covino et al. 2003a;
Rossi et al. 2002). In this case the break time can be measured also
from the polarization points, with a result that is consistent with
the limits from the light-curve. A moderately ordered magnetic field
is also required. A fit with the same model expanding in a wind
environment does not yield an acceptable fit. The best fit is shown
with a thick solid line in Figs.~\ref{fig:fit} and~\ref{fig:fit2}.

The best fit of the whole set is obtained for an MHJ jet, irrespective
of the environment structure. The ISM and Wind fits are shown with a
thick dash-dot-dot-dot and long-dashed lines, respectively, in
Figs.~\ref{fig:fit} and~\ref{fig:fit2}. Even though the fit is good,
it is not possible to independently constrain the parameters, since
the covariances are strong and allow one to find always a minimum of
the $\chi^2$ within 1 from the absolute minimum.

Finally, the MSJ model yields a reasonable fit, even though not
formally acceptable. The fit is shown with a thin dash-dot line in
Figs.~\ref{fig:fit} and~\ref{fig:fit2}. Taking into account the
oversimplification with which this model has been computed, we do not
consider the lack of a formally adequate $\chi^2$ value a condition
strong enough to reject the model.

\section{Summary and Discussion}

We have presented a comprehensive modelling of the polarization curve
of GRB~020813. This burst is particularly suited for polarization
studies since it has an extremely smooth light curve (Gorosabel et
al. 2003; Laursen \& Stanek 2003). This is an important parameter,
since any complexity in the light curve is likely associated with the
breaking of the fireball symmetry, introducing a random fluctuation in
polarization that cannot be predicted {\it a priori} by any model
(Granot \& K\"onigl 2003; Lazzati et al. 2003). The polarization curve
of GRB~020813 (Gorosabel et al. 2003) is one of the most extended
published to date\footnote{A more extensive polarization covering has
been performed on GRB~030329 (Greiner et al. 2003). This burst,
however, has a complex light curve which prevents the comparison of
the polarization with models, as for the case of GRB~021004 (Lazzati
et al. 2003).}, certainly the most complete associated to a smooth
light curve, and is characterised by a constant position angle and a
smoothly decreasing degree of polarization. Importantly, the data
encompass the break time of the light-curve, which is a critical time
where different models make markedly different predictions (Rossi et
al. 2004).

We first discuss the magnetic patch model (Gruzinov \& Waxman 1999),
in which polarization is due to the non-perfect cancellation of highly
polarized radiation coming from a large number of ordered magnetic
field domains independent from each other. This model predicts a
strong flickering of the position angle which is not observed and can
therefore be rejected based on our dataset (see also Barth et
al. 2003 and, for the case of GRB~030329, Greiner et al. 2003).

We then model the polarization curve according to the predictions of
several different models. We find that a homogeneous jet in which the
magnetic field is shock generated (Ghisellini \& Lazzati 1999; Sari
1999) cannot fit the data, since neither a minimum of polarization nor
a rotation of the position angle are present in the data. On the other
hand a structured model, in which the core of the jet is more
energetic than its wings, can successfully reproduce the data, and
predicts a jet break-time in agreement with what measured from the
light-curve. We also compute models for a magnetised jet, in which the
magnetic field has a non negligible toroidal component. We find that
the best fit is obtained by a homogeneous magnetised jet, even though
the dataset is not extensive enough to meaningfully constrain its
parameters. We also consider a force-free magnetic bubble (Lyutikov et
al. 2003), which is analogous to a structured jet but for the presence
of a toroidal magnetic field. The fit in this case is not formally
successful, but we cannot rule out the model since the details of how
the toroidal magnetic component mixes with the shock generated one
have not been deeply investigated.

All the considered fits require a non fully ordered magnetic field,
even though the degree of order is always substantial ($>40\%$). This
is in disagreement with what is found in the prompt emission of
GRB~021206 where a fully aligned magnetic field has to be considered
in order to reproduce the observational constraints\footnote{Note
however that the result has been heavily criticised on data analysis
grounds (Rutledge \& Fox 2004) and any implication should therefore be
taken with caution.}  (Coburn \& Boggs 2003; Lyutikov et al. 2003;
Granot 2003; but see also Nakar et al. 2003 and Lazzati et
al. 2004). This indicates that the magnetic field geometry is
different in the two epochs and seems to confirm the idea that the
material responsible for the prompt emission is not the same one that
produces afterglow photons. Our best fit model requires however a
mixing of the two, or at least of their electromagnetic components.

Unfortunately the quality of the data is not good enough to allow
us to constrain the models and/or the assumption on which they are
based any further. Robustly and independently of the model assumptions
we can conclude that: i) the magnetic field responsible for afterglow
emission is not simply random since a net polarization signal is
produced; ii) the difference between the magnetic field components
parallel and perpendicular to the shock front is small but non
negligible, but it is not possible to understand which of the two
dominates over the other; iii) a standard homogeneous jet with shock
generated field can be ruled out, since no angle rotation is detected
between observations before and after the jet break\footnote{\bf A
constant position angle does not imply necessarily an inhomogeneous
jet with the structure adopted in this paper. Any jet with a brighter
core and dimmer wings will produce qualitatively similar polarization
curves.}. On the theoretical side, we show that the early time
behaviour of polarization is crucial to understand the magnetic field
configuration: a large scale field will produce large polarization at
early time, while a small-scale field will produce polarization only
within a couple of decades in time from $t\sim{}t_{\rm{jet}}$.

We conclude by commenting on how future observations may help to gain
a deeper insight on the structure of GRB jets and their magnetic
field. As shown in Fig.~\ref{fig:fit2}, the various models differ
substantially at early times. In particular, magnetised jets predict
high polarization at early times, while unmagnetised models predict
null polarization. Early polarization measurements of afterglows with
smooth light curves will therefore be fundamental to pin down these
important jet parameters.

We shall also conclude with a word of caution. Dense sampling of the
polarization curve of GRB~030329 (Greiner et al. 2003) has revealed
that the polarization curve associated to a complex light curve
afterglow can be much more complex than what predicted by any of the
models we discussed here, and that fluctuations in the light and
polarization curves may not be strictly correlated. In the case of
GRB~020813 the light and polarization curve sampling is not as dense,
and therefore even if all data are consistent with a smooth evolution,
short time-scale variability in both the light and polarization curve
may have been missed\footnote{However Lazzati (2004) showed that a
degree of fluctuation as large as that in GRB~030329 would have been
easily detected in GRB~020813.}. We therefore strongly recommend an
adequate sampling of GRB afterglow light curves in polarimetric mode
in order to allow for a more robust comparison of the data with
models.

\begin{acknowledgements}
We thank Aaron Barth for valuable information on the reduction of the
GRB~020813 spectropolarimetric data and Maxim Lyutikov for insightful
discussions on the role of magnetised GRB jets in afterglow
polarization. DL acknowledges support from the PPARC postdoctoral
fellowship PPA/P/S/2001/00268. The authors acknowledge benefits from
collaborating within the Research Training Network ``Gamma-Ray Bursts:
An Enigma and a Tool'', funded by the EU under contract
HPRN-CT-2002-00294.
\end{acknowledgements}

\end{document}